\documentclass[twocolumn,showpacs,prb]{revtex4-1}
\usepackage{bm}
\usepackage{graphicx}
\usepackage{epsfig}
\usepackage{amsmath} 
\usepackage{color}
\usepackage{ulem}
\usepackage{amssymb}
\usepackage[colorlinks = true]{hyperref}

\renewcommand{\i}{\mathrm{i}}
\newcommand{\eps}{\varepsilon}
\newcommand{\e}{\mathrm{e}}

\newcommand{\ket}[1]{\left| #1 \right\rangle}
\newcommand{\B}{\mathcal B}
\newcommand{\D}{\mathcal D}
\newcommand{\A}{\mathcal A}

\definecolor{darkgray}{rgb}{0.66, 0.66, 0.66}

\begin{document}

\title{Magnetic field effects on edge and bulk states in topological insulators based on HgTe/CdHgTe quantum wells with strong natural interface inversion asymmetry}

\author{M.\,V.\,Durnev}
\author{S.\,A.\,Tarasenko}

\affiliation{Ioffe Institute, 194021 St.\,Petersburg, Russia}

%

\begin{abstract}
We present a theory of the electron structure and the Zeeman effect for the helical edge states emerging in two-dimensional topological insulators based on HgTe/HgCdTe quantum wells with strong natural interface inversion asymmetry. The interface inversion asymmetry, reflecting the real atomistic structure of the quantum well, drastically modifies both bulk and edge states.
For the in-plane magnetic field, this asymmetry leads to a strong anisotropy of the edge-state effective $g$-factor which becomes dependent on the edge orientation.
The interface inversion asymmetry also couples the counter propagating edge states in the out-of-plane magnetic field leading to the opening of the gap in the edge-state spectrum by arbitrary small fields.
\end{abstract}

\maketitle 

\section{Introduction}

HgTe/CdHgTe quantum wells (QWs) of thickness above a critical value belong to the class of $\mathbb Z_2$ two-dimensional topological insulators characterized by the existence of counter propagating helical edge modes~\cite{Bernevig15122006, Konig:2007it, Gusev2011, RevModPhys.82.3045, RevModPhys.83.1057}. Continuous advance in the technology of CdHgTe-based heterostructures stimulates experimental and theoretical studies of their electronic properties related to the non-trivial band topology. The structure of the edge states responsible for the emergence of the quantum spin Hall effect was theoretically studied at zero magnetic field~\cite{RevModPhys.83.1057, Konig:2008fk, PhysRevLett.101.246807, PhysRevB.82.113307,  PhysRevB.91.035310, enaldiev2015}, in the presence of magnetic field lying in the QW plane~\cite{PhysRevB.85.045310, Cheng:2014ve}, and in magnetic field normal to the QW plane~\cite{PhysRevLett.104.166803, PhysRevB.86.075418, PhysRevB.85.125401, PhysRevB.91.235433}. Most of the theoretical papers consider simplified models which do not take into account the natural inversion asymmetry of the HgTe/CdHgTe QWs caused by their atomic structure or treat this asymmetry as a small perturbation~\cite{Konig:2008fk,PhysRevB.91.035310,PhysRevB.91.235433}.
Contrary, atomistic calculations performed recently~\cite{PhysRevB.91.081302, Nanostructures2014} have revealed very strong level repulsion in HgTe/CdHgTe QWs, mostly driven by the natural interface inversion asymmetry of the zinc-blende heterostructures, which results in a considerable modification of the ``bulk'' (two-dimensional) electron states and dispersion. The inversion asymmetry also gives rise to a giant photogalvanic response observed in HgTe/CdHgTe heterostructures~\cite{Olbrich2013, Dantscher2015}.  

In the present paper we study theoretically the electron structure of bulk and helical edge states in HgTe/HgCdTe QWs with strong natural interface inversion asymmetry in external magnetic field. We find that the interface mixing of the states results in (i) a strong anisotropy of the edge-state $g$-factor in the in-plane magnetic field and (ii) opening of the gap in the edge-state spectrum by an arbitrary small out-of-plane magnetic field. Both effects are absent in centrosymmetric continuum-medium models. We obtain analytical results for the energy spectrum and wave functions of the edge states in a semi-infinite two-dimensional structure and do numerical calculations of the spectrum of coupled edge states in a strip of a finite width.

The paper is organized as follows. In Sec.~\ref{Sec:bulk} we present the effective Hamiltonian of the system and describe the bulk energy spectrum at zero magnetic field and the structure of bulk Landau levels. In Sec.~\ref{sec:semi_inf} we study analytically the helical states in a semi-infinite system with a single edge at zero magnetic field (Sec.~\ref{sec:zeroB}), in the in-plane magnetic field (Sec.~\ref{sec:B_inplane}), and out-of-plane magnetic field (Sec.~\ref{sec:B_outplane}). Section~\ref{sec:orient} presents the 
study of the edge states and the Zeeman effect in a semi-infinite structure with an arbitrary orientation of the edge with respect to crystallographic axes. In Sec.~\ref{sec:finite} we outline the numeric procedure used to calculate the edge states in a strip of a finite width and compare the obtained numerical and analytical results. Sec.~\ref{sec:concl} summarizes the paper.

\section{Effective Hamiltonian and bulk Landau levels}\label{Sec:bulk}

We consider HgTe/HgCdTe QWs grown along the $z \parallel [001]$ axis ($D_{2d}$ point group) with a symmetric heteropotential. In the QWs of the critical thickness $d_c$, where the transition between the trivial and non-trivial topological phases occurs, and in QWs of close-to-critical thickness, the Dirac states are formed from the electron-like $|E1,\pm 1/2 \rangle$ and heavy-hole $|H1,\pm 3/2 \rangle$ subbands,~\cite{Bernevig15122006, Gerchikov1989}
\begin{eqnarray}
\label{eq:E1H1}
|E1,\pm 1/2 \rangle &=& f_1(z) |\Gamma_6, \pm 1/2 \rangle + f_4(z) |\Gamma_8, \pm 1/2 \rangle \:, \nonumber \\
|H1, \pm 3/2 \rangle &=& f_3(z) |\Gamma_8, \pm 3/2 \rangle \:,  
\end{eqnarray}
where $f_1(z)$, $f_3(z)$, and $f_4(z)$ are the envelope functions, $|\Gamma_8,\pm 1/2 \rangle$, $|\Gamma_8,\pm 3/2 \rangle$, and $|\Gamma_6,\pm 1/2 \rangle$ are the Bloch amplitudes of the $\Gamma_8$ and $\Gamma_6$ bands at the $\Gamma$ point of the Brillouin zone.

Symmetry lowering resulting from the anisotropy of the QW interfaces leads to an efficient interface coupling of the light-hole states $|\Gamma_8, \pm 1/2 \rangle$ and heavy-hole states $|\Gamma_8, \mp 3/2 \rangle$ and, hence, to coupling of the electron-like and heavy-hole subbands. This coupling leads to the level anticrossing at the interfaces and splitting of the Dirac cones~\cite{PhysRevB.91.081302}.  

The effective 4$\times$4 $\bm k$$\cdot$$\bm p$ Hamiltonian, which precisely takes into account the real spatial symmetry of the QW structure, can be constructed in the framework of the group representations theory. The effective Hamiltonian can be derived taking into account that, in the $D_{2d}$ point group, the $|E1,\pm 1/2 \rangle$ and $|H1, \mp 3/2 \rangle$ pairs transform according to the spinor representation $\Gamma_6$ while the components $k_x,k_y$ of the in-plane wave vector $\bm k$ belong to the irreducible representation $\Gamma_5$. The effective Hamiltonian to the second order in the wave vector in the $|E1,+\rangle,|H1,+\rangle,|E1,-\rangle,|H1,-\rangle$ basis has the form (see also Refs.~\onlinecite{Bernevig15122006, PhysRevB.91.081302, Konig:2008fk, Winkler20122096})
\begin{widetext}
\begin{equation}
\label{eq:H_bulk}
\mathcal H_0(k_x,k_y) =
\left( 
\begin{array}{cccc}
\delta_0 - (\B+\D)k^2 & {\rm i} \A k_+ & \beta_e k_+ & {\rm i} (\gamma + \gamma' k^2) \\
-{\rm i} \A k_- & - \delta_0 + (\B-\D)k^2 & {\rm i} (\gamma + \gamma' k^2) & \beta_h k_-\\
\beta_e k_- & -{\rm i} (\gamma + \gamma' k^2) & \delta_0 - (\B+\D)k^2 & - {\rm i} \A k_- \\
-{\rm i} (\gamma + \gamma' k^2) & \beta_h k_+ & {\rm i} \A k_+ & - \delta_0 + (\B-\D)k^2
\end{array}
\right) \: .
\end{equation}
\end{widetext}
Here, $k = |\bm k|$, $k_\pm = k_x \pm \i k_y$, $x \parallel [100]$ and $y \parallel [010]$ are the in-plane axes,
$\A$, $\B$, $\D$, $\beta_e$, $\beta_h$, $\gamma$, $\gamma'$, and {$\delta_0$} are the structure parameters. The parameter {$\delta_0$} determines the energy gap. It can be tuned from positive to negative values by varying the QW thickness and defines whether the system is in the trivial ({$\delta_0 >0$} at negative $\B$) or non-trivial ($\delta_0 <0$, $\B < 0$) topological phase. The parameters $\beta_e$ and $\beta_h$ describe contributions to 
$\bm k$-linear splitting of the electron-like and heavy-hole subbands caused by bulk inversion asymmetry. The parameters $\gamma$ and $\gamma'$ are determined by the interface mixing strength. Atomistic calculations yield the splitting $2|\gamma| \approx 10$~meV at $\bm k =0$  for HgTe/Hg$_{0.3}$Cd$_{0.7}$Te QWs with atomically sharp interfaces~\cite{PhysRevB.91.081302}. Such a strong interface coupling of the states drastically affects the energy spectrum and cannot be treated as a small perturbation. In contrast, the parameters $\beta_e$, $\beta_h$, and $\gamma'$ lead only to corrections to the splitting at $\bm k \neq 0$. Therefore, to simplify calculations we consider the standard Bernevig-Hughes-Zhang Hamiltonian of the HgTe-based two-dimensional topological insulators~\cite{Bernevig15122006} ($\beta_e$, $\beta_h$, $\gamma' = 0$) with account for the interface coupling described by $\gamma$. For numerical calculations presented below we use parameters: $\A = 3.6$~eV$\cdot$\AA, $\B = -68$~eV$\cdot$\AA$^2$, $\D = -51$~eV$\cdot$\AA$^2$ corresponding to a 7-nm-wide HgTe/Hg$_{0.3}$Cd$_{0.7}$Te QW (the thickness is slightly above $d_c$)~\cite{Konig:2008fk}. We assume $\gamma$ to be positive and use $\gamma = 5$~meV.

\begin{figure}[b]
\includegraphics[width=0.48\textwidth]{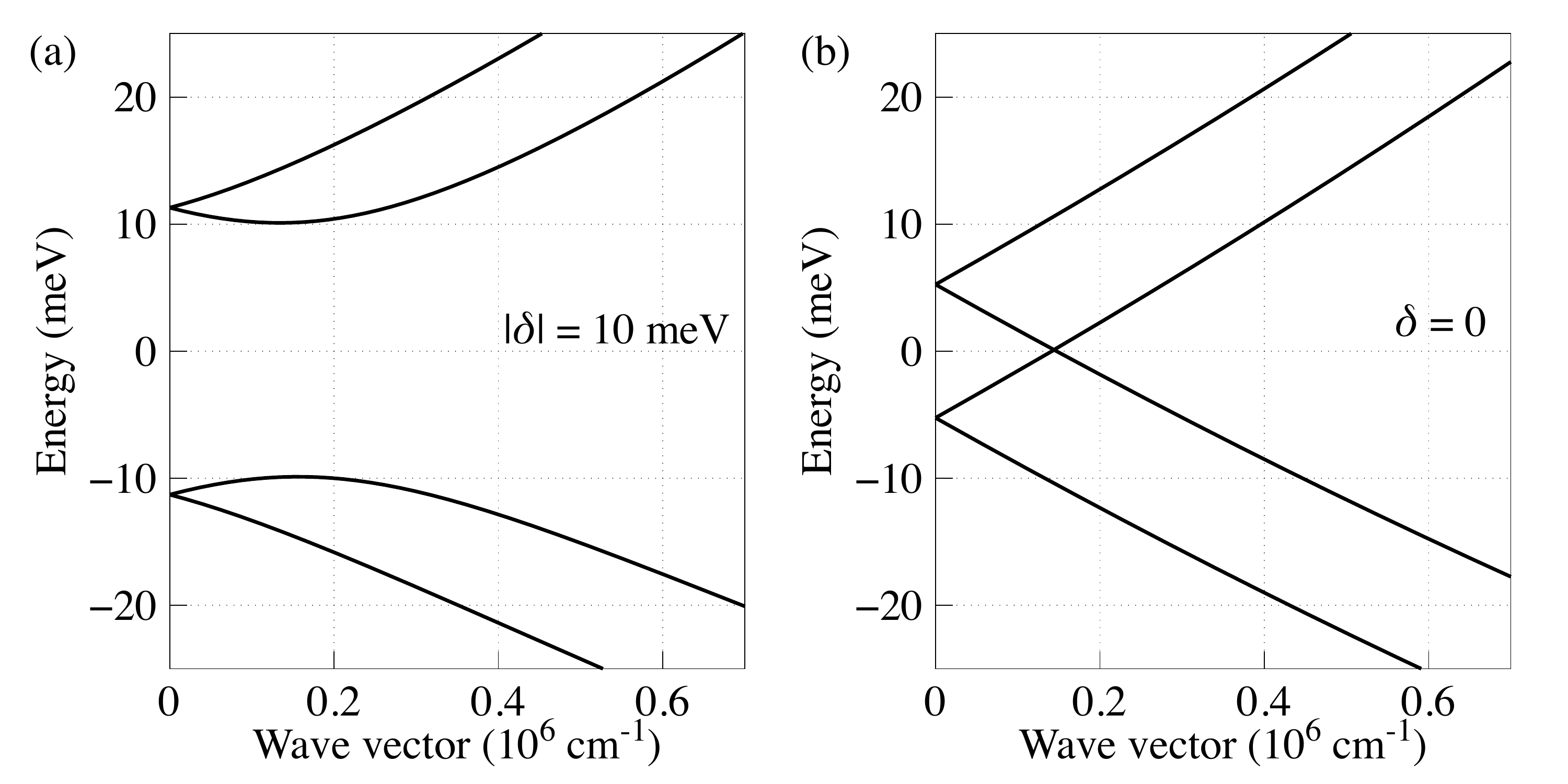}
\caption{\label{fig:zero_B}
Energy spectra of the bulk Dirac fermions in HgTe/CdHgTe QWs (a) with ${|\delta| = 10}$~meV, and (b) 
of the critical thickness, $\delta = 0$, at zero magnetic field. The level splitting at $\bm k=0$ in the panel (b) is caused by the light-hole--heavy-hole mixing.
}
\end{figure}

Diagonalization of the Hamiltonian~\eqref{eq:H_bulk} yields the energy spectrum of the bulk Dirac fermions~\cite{PhysRevB.91.081302}
\begin{equation}\label{spectrum_zeroB}
\eps(\bm k) = -\D k^2 \pm \sqrt{\left( {\delta_0} - \B k^2 \right)^2 + \left( \gamma \pm \A k \right)^2}\:.
\end{equation}
The dispersion curves given by Eq.~\eqref{spectrum_zeroB} are depicted in Fig.~\ref{fig:zero_B}. Generally, the spectrum
contains four branches with an energy gap {$2|\delta|$, where $\delta = \delta_0 - \B k_0^2$,} at the wave vector $k_0 = \gamma/\A$. For a gapless structure {($\delta = 0$)}, the spectrum consists of two massless Dirac cones shifted vertically with respect to each other by 2$\gamma$.

External magnetic field {$\bm B$} is included in the $\bm k$$\cdot$$\bm p$ theory by the Peierls substitution $\bm k \rightarrow \hat{\bm k} - {(e/c\hbar)}\bm A$ in the zero-field Hamiltonian~\eqref{eq:H_bulk} and by adding the Zeeman Hamiltonian
\begin{equation}
\label{eq:H_Bperp}
\mathcal H_{Z} = 
\frac{\mu_B}{2} \left(
\begin{array}{cccc}
g_{e}^\perp B_z & 0 & g_e^\parallel B_- & 0 \\
0 & g_{h}^\perp B_z & 0 & g_h^\parallel B_+ \\
g_e^\parallel B_+ & 0 & - g_{e}^\perp B_z & 0 \\
0 & g_h^\parallel B_- & 0 & - g_{h}^\perp B_z
\end{array}
\right) \:,
\end{equation}
where $\hat{\bm k} = - \i \nabla$, $e$ is the electron charge, $\bm A$ is the vector potential of the magnetic field, $\bm B = \nabla \times \bm A$, $\mu_B$ is the Bohr magneton, $g_e^\parallel$, $g_e^\perp$, $g_h^\parallel$, and $g_h^\perp$ are the contributions to the
$g$-factors of the $|E1 \rangle$ and $|H1 \rangle$ subbands stemming from the bare electron $g$-factor and interaction with remote electron and hole subbands, and $B_{\pm} = B_x \pm i B_y$. 
The coupling of the $|E1 \rangle$ and $|H1 \rangle$ states by the out-of-plane magnetic
field is exactly taken into account in the Hamiltonian $\mathcal H_0[\hat{\bm k} - (e/c\hbar)\bm A]$. 
The coupling of these bands by the in-plane magnetic field, which occurs in QWs of the $D_{2d}$ symmetry, is small since it requires the consideration of inversion symmetry breaking, and thus, is neglected in the Hamiltonian~\eqref{eq:H_Bperp}. We also note that $g_h^\parallel$ is expected to be small compared to $g_e^\parallel$~\cite{Mar99}. For numerical calculations we use $g_e^\parallel = -20$, $g_e^\perp = 22$, $g_h^\perp = -1$, and $g_h^\parallel = 0$~\cite{Konig:2008fk}.
  
The structure of the bulk Landau levels in the perpendicular magnetic field $\bm B = (0,0,B_z)$ can be readily found by solving the Schr\"{o}dinger equation $\mathcal H \Psi = \varepsilon \Psi$ 
with the Hamiltonian
\begin{equation}
\label{eq:H_Bfield}
\mathcal H = \mathcal H_0[\hat{\bm k} - (e/c\hbar)\bm A] + \mathcal H_{Z} \:.
\end{equation}
We take the vector potential in the Landau gauge $\bm A = (0, B_z x, 0)$ and following Refs.~\onlinecite{Rashba1960, Konig:2008fk, PhysRevB.86.075418} solve the  Schr\"{o}dinger equation by decomposing the four-component wave function $\Psi$ in a series of the Landau level functions $\phi_{n,k_y}$ 
\begin{equation}
\Psi = \sum_{n \geq 0} \left( 
\begin{array}{c}
a_n \\ b_n \\ c_n \\ d_n
\end{array}
\right) \phi_{n,k_y} \:,
\end{equation}
where $n$ and $k_y$ are the quantum numbers, and $a_n$, $b_n$, $c_n$, and $d_n$ are coefficients. 

\begin{figure}[t]
\includegraphics[width=0.45\textwidth]{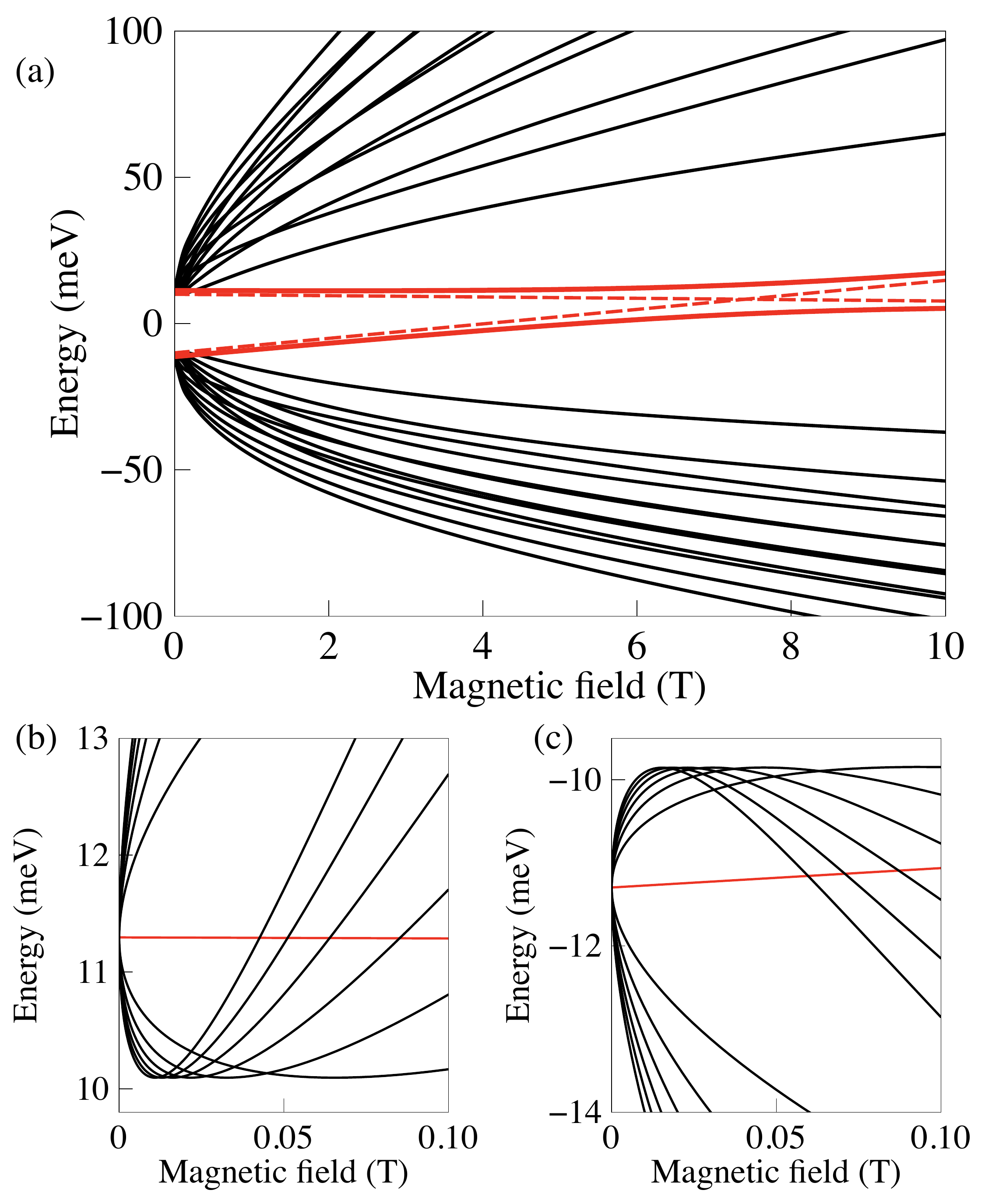}
\caption{\label{fig:LL1}
Bulk Landau levels in a HgTe/HgCdTe QW with $\delta = -10$~meV. Red curves show the magnetic field dependence of the zero Landau levels.
Dashed red lines in panel (a) show the positions of the zero Landau levels in the absence of the light-hole--heavy-hole mixing ($\gamma = 0$). Panels (b) and (c) show the structure of the Landau levels at small magnetic fields.
}
\end{figure}

\begin{figure}[t]
\includegraphics[width=0.45\textwidth]{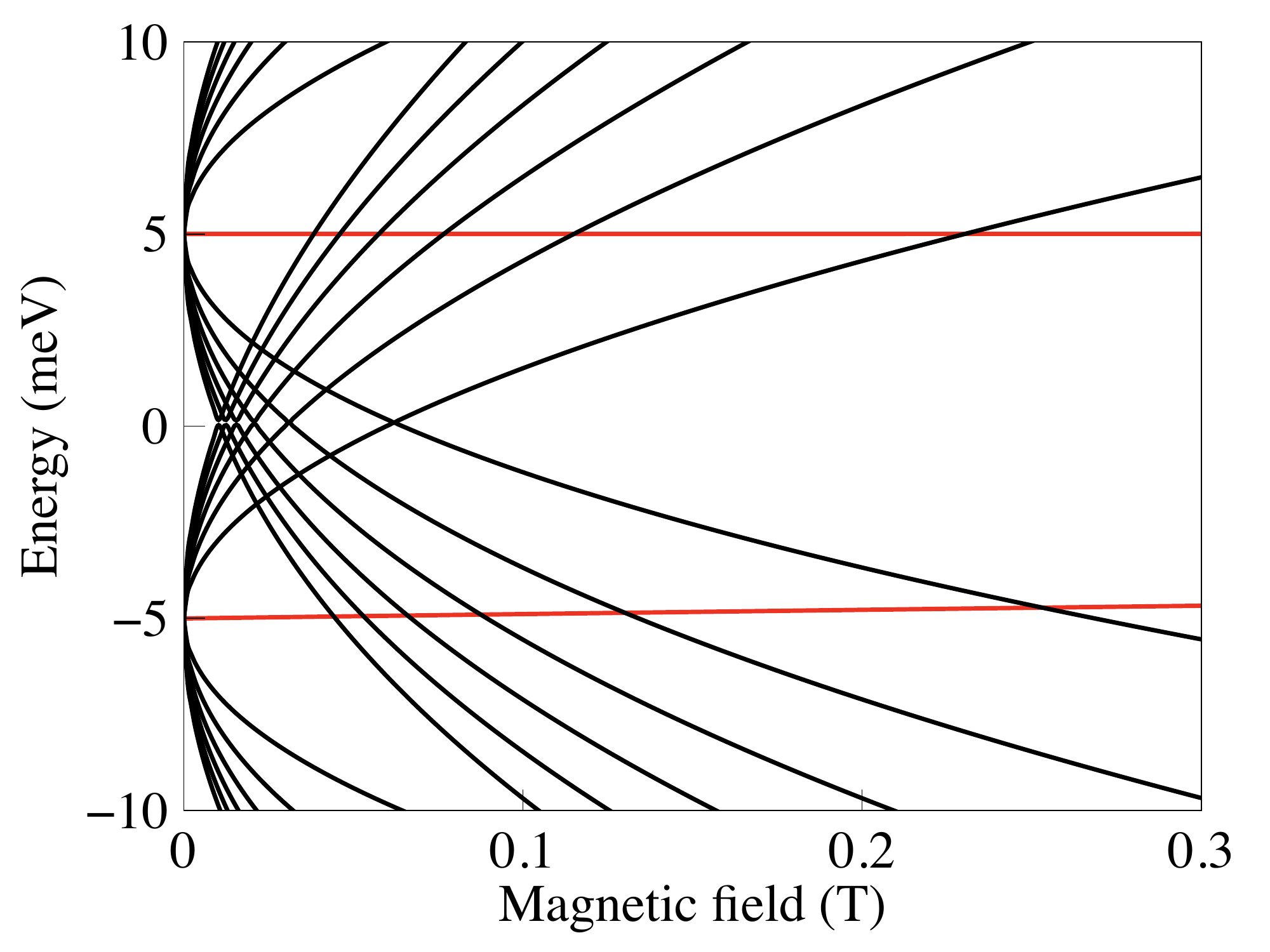}
\caption{\label{fig:LL2}
Bulk Landau levels in a HgTe/HgCdTe QW of the critical thickness, $\delta = 0$~meV. Red lines present the dispersion of ``zero'' Landau levels.
}
\end{figure}

Figure~\ref{fig:LL1} shows the calculated energy spectrum for a HgTe/CdHgTe QW in the topologically non-trivial phase.
Landau levels can be divided into two groups. The first group comprises two ``zero'' modes corresponding to $n = 0$. These modes are formed by the $\ket{E1,+}$ and $\ket{H1,-}$ subbands only and are decoupled from other Landau levels with $n \geq 1$. The energies of the zero modes are given by
\begin{multline}
\label{eq:zero_modes}
\eps_0^{(\pm)} = - \frac{\D}{l_B^2} + \frac14 (g_e^\perp - g_h^\perp) \mu_B B_z \\
 \pm \sqrt{\left[ \delta_0 - \frac{\B}{l_B^2} + \frac14 (g_e^\perp + g_h^\perp) \mu_B B_z \right]^2 + \gamma^2} \:,
\end{multline}
where $l_B = \sqrt{c\hbar/|e| B_z}$ is the magnetic length. In the absence of interface mixing ($\gamma = 0$),
the zero Landau levels cross at the critical field $B_c = \delta_0 / [|e|\B/(c\hbar) - \mu_B(g_e^\perp + g_h^\perp)/4]$~\cite{Konig:2008fk, PhysRevB.86.075418}. The mixing of these states due to interface inversion asymmetry (or bulk inversion asymmetry~\cite{PhysRevB.83.115307,PhysRevB.86.205420}) leads to coupling of the zero modes and anticrossing with the gap $2\gamma$ at $B_z = B_c$, see Fig.~\ref{fig:LL1}.

Landau levels that belong to the second group are described by the wave functions $\Psi = (a_n \ket{n}, b_{n-1} \ket{n-1}, c_{n-1} \ket{n-1}, d_n \ket{n})^T$ with $n \geq 1$. The corresponding energies are the roots of a fourth-degree polynomial. For $g_e^\perp = g_h^\perp = 0$ and $\B = \D = 0$, the energies can be found analytically and have the form
\begin{eqnarray}
\eps_n^{(1,4)} &=& \mp \sqrt{\delta_0^2 + \left( \gamma + \A \sqrt{n} /l_B \right)^2}\:, \\
\eps_n^{(2,3)} &=& \mp \sqrt{\delta_0^2 + \left( \gamma - \A \sqrt{n} / l_B \right)^2}\:. \nonumber
\end{eqnarray}

The structure of bulk Landau levels in the QW of the critical thickness is shown in Fig.~\ref{fig:LL2}. In this case,
each of the Dirac cones [Fig.~\ref{fig:zero_B}(b)] forms a fan of Landau levels.

\section{Edge states in a semi-infinite system}\label{sec:semi_inf}
\subsection{Zero magnetic field}\label{sec:zeroB}

We consider now a semi-infinite QW structure in the halfspace $x \geq 0$. In the topologically non-trivial phase $\delta < 0$, the structure supports edge states, which exponentially decay at $x \to +\infty$. We find the energy spectrum and wave functions of the edge states by solving the Schr\"{o}dinger equation 
\begin{equation}
\label{eq:Schr}
\mathcal H (\hat{k}_x, \hat{k}_y) \Psi(x,y) = \eps \Psi(x,y) \:,
\end{equation}
with the open boundary conditions at the edge, $\Psi(0,y) = 0$. A more general form of boundary conditions was studied in Ref.~\onlinecite{enaldiev2015}.

Taking into account the translation invariance along the $y$ direction we can present the wave function of the edge state in the form
\begin{equation}
\label{eq:ansatz1}
\Psi(x,y) = \frac{ \e^{\i k_y y}}{\sqrt{L}} \sum \limits_{j=1}^{8} c_j \e^{-\lambda_j x} \xi_j \:,
\end{equation}
where $k_y$ is the wave vector along the edge, $L$ is the normalization length, $c_j$ are the coefficients to be determined from the boundary conditions, $\lambda_j$ are the complex-valued reciprocal lengths, and $\xi_j$ are the position-independent
normalized four-component columns. We note that all components of $\xi_j$ are generally nonzero since the Hamiltonian~\eqref{eq:H_bulk} contains off-diagonal blocks $\propto \gamma$. This is in contrast to centrosymmetric models ($\gamma = 0$) with the 
decoupled lower and upper blocks of the Hamiltonian.

For a given wave vector $k_y$, the columns $\xi_j$ and the relation between the reciprocal lengths $\lambda_j$ and the energy $\eps$
are found from the matrix equation
\begin{equation}
\label{eq:Hlambda}
\mathcal H(\i \lambda, k_y) \xi = \eps \xi \:.
\end{equation}

First, we calculate the wave functions at $k_y=0$. We consider the case of ``electron-hole'' symmetry, i.e., $\D=0$ in the Hamiltonian~\eqref{eq:H_bulk}, so that the edge states at $k_y = 0$ have the energy $\eps = 0$. A more general case of $\D \neq 0$ will be analyzed
in Sec.~\ref{sec:finite} in numeric simulations of the electron states in a strip of a finite width.
For $\eps = 0$, the right-hand side of Eq.~\eqref{eq:Hlambda} vanishes and non-trivial solutions for $\xi$ exist for a set of $\lambda$ which satisfy the equation $\mathrm{det}~ \mathcal H (\mathrm{i} \lambda, 0)  = 0$. This equation yields eight reciprocal lengths $\lambda_j$ which are pairwise related to each other by complex conjugation
\begin{align}\label{eq:lambdas}
\lambda_1 = \lambda_2^* = - \lambda_5 = -\lambda_6^* = -\frac{\A + \sqrt{\A^2 - 4\B(\delta + \mathrm{i} \gamma)}}{2\B}\:, \nonumber \\
\lambda_3 = \lambda_4^* = - \lambda_7 = -\lambda_8^* = -\frac{\A - \sqrt{\A^2 - 4\B(\delta + \mathrm{i} \gamma)}}{2\B}\:.
\end{align}
The corresponding normalized null-space vectors of the matrices $\mathcal H(\i \lambda_j, 0)$ have the form
\begin{align}
\xi_1 = \xi_3 = \frac12 \left( 
\begin{array}{c}
1 \\
-1 \\
1\\ 
1
\end{array} \right)\:,\:\:
\xi_2 = \xi_4 = \frac12 \left( 
\begin{array}{c}
-1 \\
1 \\
1\\ 
1
\end{array} \right)\:, \nonumber \\
\xi_5 = \xi_7 = \frac12 \left( 
\begin{array}{c}
1 \\
1 \\
-1\\ 
1
\end{array} \right)\:,\:\:
\xi_6 = \xi_8 = \frac12 \left( 
\begin{array}{c}
1 \\
1 \\
1 \\ 
-1
\end{array} \right)\:.
\end{align}
The wave functions of the edge states are given by Eq.~\eqref{eq:ansatz1}. To satisfy the boundary condition $ \sum_j c_j \xi_j =0 $ and
the wave function decay at $x \to +\infty$ (implying $\mathrm{Re}~\lambda > 0$) one has to set $c_1 = -c_3$, $c_2 = -c_4$, and 
$c_5 = c_6 = c_7 = c_8 = 0$. Finally, the wave functions of the edge states are two-fold degenerate at  $k_y = 0$ and can be chosen in the form
\begin{eqnarray}\label{eq:decomp2}
\Psi_1(x,y) = C \frac{ \e^{\i k_y y}}{\sqrt{L}} \left( \e^{-\lambda_1 x} - \e^{-\lambda_3 x} \right) \xi_1 \:, \nonumber \\
\Psi_2(x,y) = C \frac{ \e^{\i k_y y}}{\sqrt{L}} \left( \e^{-\lambda_1^* x} - \e^{-\lambda_3^* x} \right) \xi_2 \:,
\end{eqnarray}
where 
\[
C = \left[ \frac{1}{2 \mathrm{Re} \lambda_1 } + \frac{1}{2 \mathrm{Re} \lambda_3} - 2 \mathrm{Re}\left\{ \frac{1}{\lambda_1 + \lambda_3^*} \right\} \right]^{-1/2}
\]
is the normalization constant. We note that in topologically trivial phase ($\delta > 0$), the boundary conditions yield $c_j = 0$ for all the coefficients and, hence, no edge states emerge.

The dependence of the wave functions~\eqref{eq:decomp2} on $x$ can be further simplified by introducing the 
effective lengths $l_1 = - \B/\A$ and $l_2 = - \A/ \delta$ and the wave vector $k_0 = \gamma/\A$. For the parameters of HgTe/CdHgTe QWs and $\delta = -10\,$meV, one has $l_1 \approx 20$~\AA, $l_2 \approx 360$~\AA, and $1/k_0 \approx 600$~\AA.
Taking into account that $l_1 \ll l_2, \, 1/k_0$, we obtain
\begin{equation}
\lambda_1 \approx \frac{1}{l_1}\:,\:\:\: \lambda_3 \approx \frac{1}{l_2} - i k_0 \:.
\end{equation}
Note that the wave vector $k_0$ corresponds to the position of the bulk energy gap in the two-dimensional Brillouin zone, see Sec.~\ref{Sec:bulk}.

To obtain the energy spectrum at $k_y \neq 0$ we construct the effective 2$\times$2 Hamiltonian by projecting the Hamiltonian~\eqref{eq:H_bulk} onto the basis functions $\Psi_1$ and $\Psi_2$. To the first order in $k_y$, the effective Hamiltonian of the edge states reads
\begin{equation}\label{eq:H_edge}
\mathcal H_{{\rm edge}}^{(\Psi)} (k_y) = - \frac{A \delta k_y}{\delta^2 + \gamma^2} \left( 
\begin{array}{cc}
0 & \delta + \i \gamma \\
\delta - \i \gamma & 0 
\end{array}
\right) .
\end{equation}
The Hamiltonian~\eqref{eq:H_edge} is diagonalized by the unitary transformation
\begin{equation}
\label{edge_wfs}
\left( \begin{array}{c}
\Phi_1 \\
\Phi_2
\end{array} \right) = 
\frac{1}{\sqrt{2}}\left( \begin{array}{cc}
\e^{-\i \varphi/2} & - \e^{\i \varphi/2} \\
\e^{-\i \varphi/2} & \e^{\i \varphi/2} \\
\end{array} \right)
\left( \begin{array}{c}
\Psi_1 \\
\Psi_2
\end{array} \right) ,
\end{equation}
where $\varphi = \arctan (-\gamma/\delta)$. In the basis of the functions $\Phi_1$ and $\Phi_2$, it has the form
\begin{equation}\label{eq:H_edge2}
\mathcal H_{\mathrm{edge}}^{(\Phi)} (k_y) = v \hbar k_y \sigma_z \:,
\end{equation}
where $\sigma_z$ is the Pauli matrix and
\begin{equation}\label{eq:velocity}
v = - \frac{A \delta /\hbar }{\sqrt{\delta^2 + \gamma^2}} \:.
\end{equation}
The $v$ parameter describes the group velocity of the edge states. In contrast to models neglecting the light-hole--heavy-hole mixing, the velocity $v$ at $\gamma \neq 0$ depends on the QW thickness. At $|\delta| \ll \gamma$, the velocity tends to zero, and the dispersion of the edge states vanishes.

The wave functions $\Phi_{1, 2}$ can be written in an equivalent form
\begin{eqnarray}
\label{eq:Psi_ab}
\Phi_1 &=& \frac{ \e^{\i k_y y}}{\sqrt{L}}  \left[ a(x) \frac{\xi_1 - \xi_2}{\sqrt{2}} - \i b(x) \frac{\xi_1 + \xi_2}{\sqrt{2}} \right]\:, \nonumber \\
\Phi_2 &=& \frac{ \e^{\i k_y y}}{\sqrt{L}} \left[ a(x) \frac{\xi_1 + \xi_2}{\sqrt{2}} - \i b(x) \frac{\xi_1 - \xi_2}{\sqrt{2}} \right]\:,
\end{eqnarray}
where $a(x)$ and $b(x)$ are the real functions
\begin{align}
\label{eq:ab}
a(x) =  \sqrt{\frac{2}{l_2}} \left[ \e^{-x/l_1} \cos \frac{\varphi}{2} - \e^{-x/l_2} \cos \left( k_0 x - \frac{\varphi}{2} \right) \right]\:, \nonumber \\
b(x) = \sqrt{\frac{2}{l_2}} \left[ \e^{-x/l_1} \sin \frac{\varphi}{2} + \e^{-x/l_2} \sin \left( k_0 x - \frac{\varphi}{2} \right) \right]\:.
\end{align}
Note that $b(x)$ is nonzero only at $\gamma \neq 0$.

\begin{figure}[t]
\includegraphics[width=0.9\linewidth]{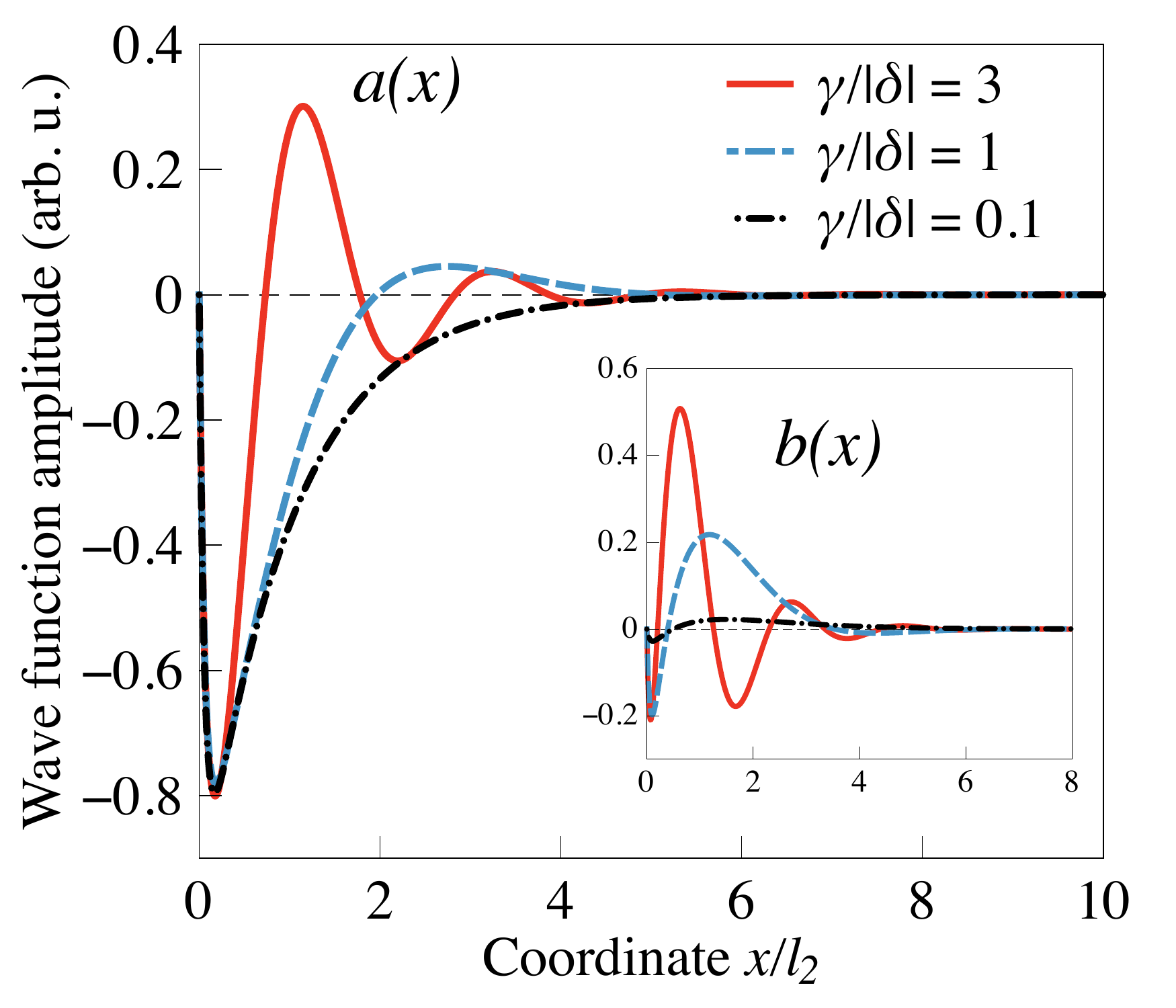}
\caption{\label{fig:fig1} Functions $a(x)$ and $b(x)$ which determine the spatial structure of the edge-state wave functions
$\Phi_{1, 2}$, see Eq.~\eqref{eq:Psi_ab}, for different values of $k_0 l_2 = \gamma/|\delta|$.
}
\end{figure}

The amplitudes $a(x)$ and $b(x)$ are depicted in Fig.~\ref{fig:fig1}. At $x \gg l_1$, $a(x)$ and $b(x)$ (i) oscillate with 
the wave length $2\pi/k_0$ and (ii) decay with the characteristic length $l_2$. The number of oscillations within the decay length is given by the dimensionless parameter $k_0 l_2 = -\gamma/\delta$ which is large for $|\delta| \ll \gamma$.

\subsection{In-plane magnetic field effect on edge states} \label{sec:B_inplane}

We consider now the effect of in-plane magnetic field $\bm B = (B_x, B_y, 0)$ on the edge states in a semi-infinite structure with the edge perpendicular to the $x \parallel [100]$ axis. The in-plane field couples the electron as well as the hole spin states,
see the Zeeman Hamiltonian~\eqref{eq:H_Bperp}. By projecting $\mathcal H_Z$ onto the edge states $\Phi_{1,2}$, we obtain the  effective Zeeman Hamiltonian 
\begin{equation}
\label{eq:Hedge_B}
{\mathcal H_{B,\rm edge}^{(\Phi)}} = \frac12 \mu_B \left( g_{xx} \sigma_x B_x + g_{yy} \sigma_y B_y \right)\:,
\end{equation}
with the components of the effective $g$-factor tensor of the edge states given by
\begin{eqnarray}
\label{eq:gfactors}
g_{xx} &=& \frac12(g_e^\parallel - g_h^\parallel)\:,\\ 
g_{yy} &=& \frac12 (g_e^\parallel + g_h^\parallel) \frac{-\delta}{\sqrt{\delta^2 + \gamma^2}} \:. \nonumber
\end{eqnarray}
In the Hamiltonian~\eqref{eq:Hedge_B}, we neglected the term $g_{zy} \sigma_z B_y$ since the off-diagonal component $g_{zy}$ of 
the edge $g$-factor tensor is proportional to small parameters $l_1/l_2$ and $k_0 l_1$.

Equations~\eqref{eq:Hedge_B}, \eqref{eq:gfactors} present one of the main results of our work: The light-hole--heavy-hole mixing provided by the D$_{2d}$ point-group symmetry of the QW results in a strong anisotropy of the in-plane Zeeman effect for the edge states. The effective $g$-factor for the magnetic field pointing along the edge, $\bm B \parallel y \parallel [010]$, is reduced by the factor $1/\sqrt{1+(\gamma/\delta)^2}$ which is determined by the ratio between the energy of the light-hole--heavy-hole mixing and the bulk band gap.   

Magnetic field opens the gap in the energy spectrum of the edge states
\begin{equation}
\label{eq:gap2}
\eps_{\rm gap} = \mu_B B \sqrt{ g_{xx}^2 \cos^2 \alpha + g_{yy}^2 \sin^2 \alpha} \:,
\end{equation}
where $\alpha = \arctan(B_y/B_x)$ is the angle between the in-plane vector $\bm B$ and the edge normal. Due to the $g$-factor anisotropy given by Eqs.~\eqref{eq:gfactors}, the gap depends on the magnetic field orientation.

\subsection{Out-of-plane magnetic field effect on edge states} \label{sec:B_outplane}

Magnetic field perpendicular to the QW plane $\bm B = (0,0,B_z)$ affects the edge states via the coupling of the electron-like and 
heavy-hole subbands included in the Hamiltonian $\mathcal H_0[\hat{\bm k} - (e/c\hbar)\bm A]$ and the extra
Zeeman term~\eqref{eq:H_Bperp}. Although magnetic field breaks the time reversal symmetry and, consequently, may destroy topological protection of the edge states, it was shown that in centrosymmetric models magnetic field $B_z < B_c$ does not open the gap in the spectrum of the edge states preserving their helical
structure~\cite{PhysRevB.86.075418, PhysRevB.91.235433}. The gap is opened in high enough fields only ($B_z > B_c$), when the system is in the quantum Hall effect regime. The interface mixing in real QWs qualitatively changes the behavior of the edge states in out-of-plane magnetic field and leads to the emergence of the gap at arbitrary small fields. 

To analyze the edge-state spectrum in the out-of-plane magnetic field we take the vector potential in the form $\bm A = (0, B_z (x+x_c), 0)$, where $x_c$ is a constant. Different $x_c$ correspond to different gauges of the magnetic field. By projecting $\mathcal H_0[\hat{\bm k} - (e/c\hbar)\bm A] + \mathcal H_{Z}$ onto the edge states $\Phi_{1,2}$, we obtain the effective Zeeman Hamiltonian
\begin{equation}
\label{eq:Hedge_B_out}
\mathcal H_{B,\rm edge}^{(\Phi)} = \frac12 \mu_B \left( g_{yz} \sigma_y B_z + g_{zz} \sigma_z B_z \right)\:,
\end{equation}
where the components of the $g$-factor tensor at $k_y=0$, in the leading order in $l_1/l_2$ and $k_0 l_1$, are given by
\begin{eqnarray}
\label{eq:gfactor_Bz}
g_{zz} &=& \frac{g_e^\perp + g_h^\perp}{2}  \frac{-\delta}{\sqrt{\delta^2 + \gamma^2}} + \frac{2m_0 {\cal A}^2}{\hbar^2}  \frac{\delta^2 +2(\delta^2 + \gamma^2)(x_c/l_2)}{\left( \delta^2 + \gamma^2 \right)^{3/2}} , \nonumber \\ 
g_{yz} &=& \frac{2 m_0 {\cal A}^2}{\hbar^2}  \frac{- \delta\gamma}{\left( \delta^2 + \gamma^2 \right)^{3/2}} \:,
\end{eqnarray}
and $m_0$ is the free electron mass. Both $g_{zz}$ and $g_{yz}$ components have large orbital contributions originating from the term $\mathcal H_0[\hat{\bm k} - (e/c\hbar)\bm A]$. 

From the Hamiltonians~\eqref{eq:H_edge2} and~\eqref{eq:Hedge_B_out} we deduce that the diagonal component $g_{zz}$ leads only to a shift of the energy spectrum along $k_y$ without opening the gap. The shift depends on the magnetic field gauge and, for a single edge, can be excluded by a proper choice of the coordinate frame. In contrast, the term $\propto g_{yz} \sigma_y B_z$ is gauge independent. It couples edge states with the opposite pseudospin projections and opens the gap
\begin{equation}
\label{eq:gap_Bz}
\eps_{\rm gap} = \frac{2 m_0 {\cal A}^2}{\hbar^2}  \frac{ \mu_B |\delta \gamma B_z| }{\left( \delta^2 + \gamma^2 \right)^{3/2}} \:.
\end{equation}
The gap in the edge-state spectrum emerges due to the lack of space inversion asymmetry in the QW and is a non-monotonic function
of the bulk band gap $2|\delta|$.

\subsection{Edge of arbitrary crystallographic orientation} \label{sec:orient}

In above subsections we studied a semi-infinite system with the edge parallel to one of the cubic axes, $y \parallel [010]$. 
Meanwhile, since the interface inversion asymmetry is associated with the certain crystallographic axes, it is natural to expect 
that the edge-state $g$-factors depend on the orientation of the edge. Now, we consider a semi-infinite structure with
the edge of an arbitrary orientation defined by the angle $\theta$ between the edge and the $[010]$ axis, see the inset in Fig.~\ref{fig:fig4}.

We introduce a new coordinate frame ($x',y',z$) rotated with respect to the frame ($x,y,z$) by the angle $\theta$ and the corresponding basis states
\begin{eqnarray}\label{eq:newstates}
|E1,\pm 1/2 \rangle' &=& |E1,\pm 1/2 \rangle {\rm e}^{\pm {\rm i} \theta /2} \:, \nonumber \\
|H1, \pm 3/2 \rangle' &=& |H1, \pm 3/2 \rangle {\rm e}^{\pm 3 {\rm i} \theta /2} \:.  
\end{eqnarray}
The effective $\bm k$$\cdot$$\bm p$ and Zeeman Hamiltonians in the $|E1,+\rangle',|H1,+\rangle',|E1,-\rangle',|H1,-\rangle'$ basis can be obtained from the Hamiltonians~\eqref{eq:H_bulk} and~\eqref{eq:H_Bperp} taking into account the wave function transformation~\eqref{eq:newstates} and the relations $k_{\pm} =  k'_{\pm} {\rm e}^{\mp {\rm i}\theta}$ and $B_{\pm} =  B'_{\pm} {\rm e}^{\mp {\rm i}\theta}$. Such a procedure shows that the rotation is equivalent to the substitution $\gamma \to \gamma \e^{-2\i\theta}$, $\gamma' \to \gamma' \e^{-2\i\theta}$, $\beta_e \to \beta_e \e^{-2\i\theta}$ and
$\beta_h \to \beta_h \e^{-2\i\theta}$ in the upper triangular part of the $\bm k$$\cdot$$\bm p$ Hamiltonian~\eqref{eq:H_bulk} and $g_h^{\parallel} \to g_h^{\parallel} \e^{-4\i\theta}$ in the upper triangular part of the Zeeman Hamiltonian~\eqref{eq:H_Bperp}. The corresponding lower triangular parts of the Hamiltonians are found from the condition of Hermitian conjugation. We note that all other contributions to the Hamiltonians~\eqref{eq:H_bulk} and~\eqref{eq:H_Bperp} possess an axial symmetry and remain unchanged under a coordinate frame rotation. As in the above consideration, we neglect the terms $\propto \beta_e$, $\beta_h$, and $\gamma'$. 

The calculation of the edge states, similar to that carried out in Subsec.~\ref{sec:zeroB}, shows that the parameters $\lambda_j$ are not affected by the rotation whereas the components of the wave functions acquire phase factors. The basis functions of the edge states, which are related to each other by time reversal, can be presented in the form
\begin{equation}
\label{eq:app_wfs}
\Phi'_1 =  \frac{ \e^{\i k_{y'} y'}}{\sqrt{2 L}}
\left[
\begin{array}{c}
a(x') \\
- a(x')  \\
- \i b(x') \e^{2 \i\theta}\\
- \i b(x') \e^{2 \i\theta}
\end{array}
\right] , \;
\Phi'_2 = \frac{ \e^{\i k_{y'} y'}}{\sqrt{2 L}}
\left[
\begin{array}{c}
-\i b(x') \e^{-2\i\theta}\\
\i b(x') \e^{-2\i\theta}\\
a(x') \\
a(x') 
\end{array}
\right] \:.
\end{equation}

The zero-field effective Hamiltonian of the edge states in the $(\Phi_1', \Phi_2')$ basis has the form
\begin{equation}
\mathcal H_{\mathrm{edge}}^{(\Phi)} (k_{y'}) = v \hbar k_{y'} \sigma_z \:,
\end{equation}
whereas the effective Zeeman Hamiltonian is given by
\begin{equation}
\label{eq:Hedge_B_theta}
\mathcal H_{B,\rm edge}^{'(\Phi)} = \frac12 \mu_B \sum \limits_{\alpha, \beta = x',y',z} g_{\alpha \beta} \sigma_\alpha B_\beta\:,
\end{equation}
with the following components of the $g$-factor tensor:
\begin{eqnarray}
\label{eq:app_gfactors}
g_{x'x'} &=& \frac12 \left( g_e^\parallel - g_h^\parallel \right) \cos^2 2\theta - \frac12 \left( g_e^\parallel + g_h^\parallel \right) \frac{\delta}{\sqrt{\delta^2+\gamma^2}} \sin^2 2\theta
, \nonumber \\
g_{y'y'} &=& \frac12 \left( g_e^\parallel - g_h^\parallel \right) \sin^2 2\theta - \frac12 \left( g_e^\parallel + g_h^\parallel \right) \frac{\delta}{\sqrt{\delta^2+\gamma^2}} \cos^2 2\theta
, \nonumber \\
g_{x'y'} &=& g_{y'x'} = \frac14 \left[ \left( g_e^\parallel - g_h^\parallel \right) + \left( g_e^\parallel + g_h^\parallel \right) \frac{\delta}{\sqrt{\delta^2 + \gamma^2}} \right] \sin 4 \theta, \nonumber \\
g_{x'z} &=& -\frac{2 m_0 {\cal A}^2}{\hbar^2} \frac{- \delta\gamma}{\left( \delta^2 + \gamma^2 \right)^{3/2}} \sin 2\theta\:, \nonumber \\
g_{y'z} &=& \frac{2 m_0 {\cal A}^2}{\hbar^2}  \frac{- \delta\gamma}{\left( \delta^2 + \gamma^2 \right)^{3/2}} \cos 2\theta \:.
\end{eqnarray}

In-plane magnetic field opens the gap in the edge-state spectrum
\begin{equation}\label{eq:gap_arb_orient}
\eps_{\rm gap} = \mu_B \sqrt{ (g_{x'x'} B_{x'} + g_{x'y'} B_{y'})^2 + (g_{y'y'} B_{y'} + g_{y'x'} B_{x'})^2 } .
\end{equation}
The dependence of the gap on magnetic field orientation at $g_h^\parallel = 0$ is given by
\begin{multline}
\label{eq:gap_theta}
\eps_{\rm gap} = \frac12 | g_e^\parallel | \mu_B B  \\
\times \sqrt{\cos^2(\alpha - 2\theta) + \frac{\delta^2}{\delta^2 + \gamma^2} \sin^2(\alpha - 2\theta)}\:,
\end{multline}
where $\alpha$ is the angle between $\bm B$ and the edge normal. The gap induced by normal magnetic field $B_z$ is given by Eq.~\eqref{eq:gap_Bz} and is independent of the edge orientation.

\section{Edge states in a strip of a finite width. Results and discussion}\label{sec:finite}

Now we present the numerical results for the energy spectrum of the edge and bulk states in the strip of a finite width $w = 1\,$$\mu$m
and compare it with the analytical theory developed in Sec.~\ref{sec:semi_inf}. We consider the strips with different crystallographic
orientations and analyze the spectrum modification in both in-plane and out-of-plane magnetic fields. To calculate the spectrum we
numerically solve the Schr\"{o}dinger equation with the Hamiltonian $\mathcal H_0[-\i \nabla - (e/\hbar c)\bm A] + \mathcal H_Z$ using the open boundary conditions $\Psi(-w/2,y) = \Psi(w/2,y) = 0$. The band structure parameters used in the calculations are listed in Sec.~\ref{Sec:bulk}. We consider HgTe/Cd$_{0.7}$Hg$_{0.3}$Te QWs in the regime of topological insulator ($\delta <0$). Varying 
the absolute value of $\delta$ allows us to simulate the QWs of different thickness. Small change of the band structure parameters upon variation of the QW thickness in the vicinity of $d_c$ (see parameterizations for different well thickness in Ref.~\onlinecite{Buttner:2011ve}) is neglected in the calculations.

\subsection{Zero magnetic field}

\begin{figure}[t]
\includegraphics[width=0.48\textwidth]{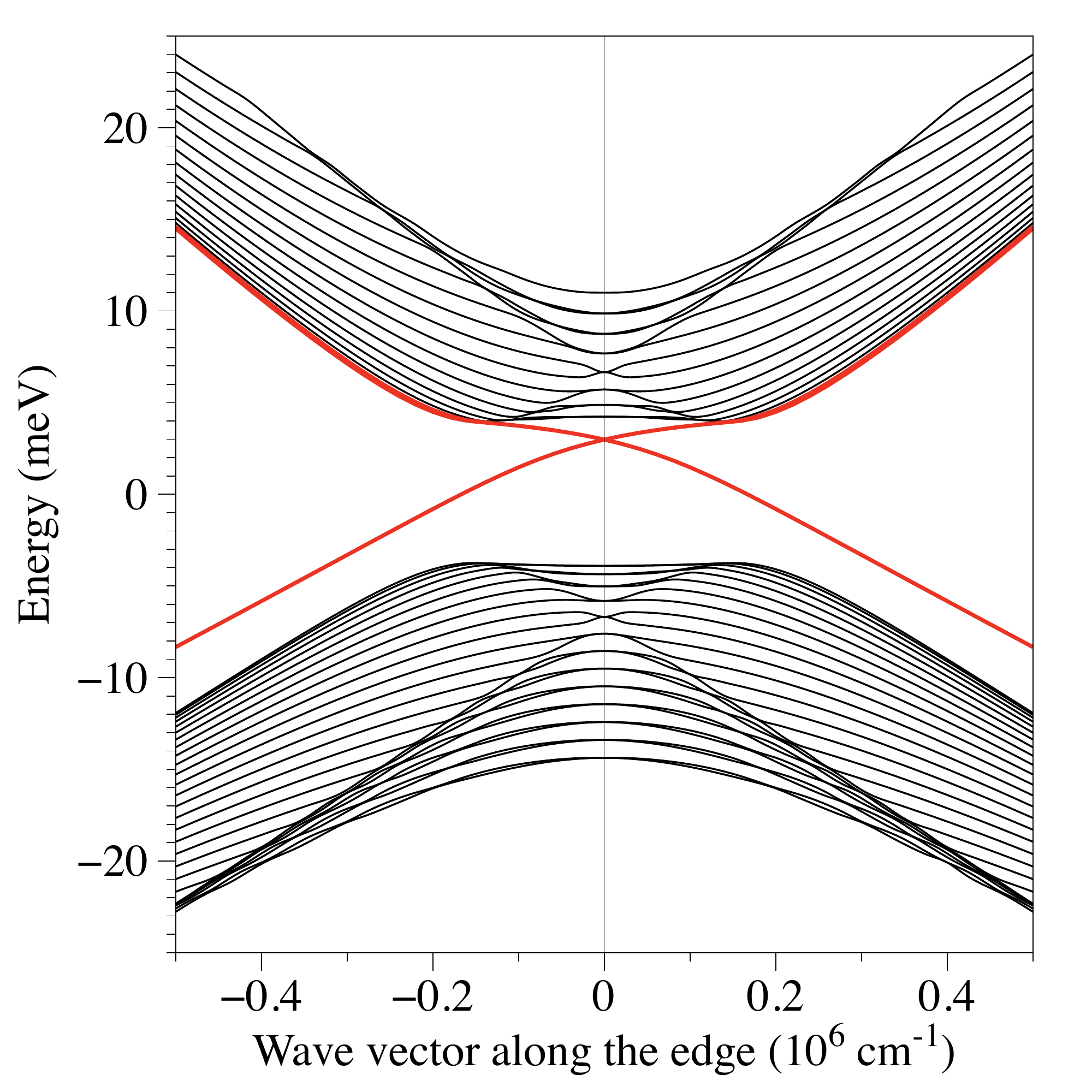}
\caption{\label{fig:fig2}
Electron energy spectrum of the strip made of HgTe/CdHgTe QW structure in the topologically non-trivial phase ($\delta = -4$~meV). The width of the strip is $1\,\mu$m.
}
\end{figure}

Figure~\ref{fig:fig2} shows the calculated zero-field energy spectrum of electron states in the strip with the bulk band gap $2|\delta| = 8$~meV. One can see that the states with linear dispersion in the vicinity of $k_y = 0$ emerge inside the band gap. The dispersion of these states is shifted from the band gap center towards positive energies due to the ``electron-hole asymmetry'', i.e. ${\cal D} \neq 0$. Each of the dispersion curves depicts two almost degenerate states corresponding to the pair of spin-polarized states localized at the two strip edges. The finite width of the strip leads to an inevitable overlap of the states localized at the spatially separated edges and, consequently, to the opening of a small gap at $k_y = 0$~\cite{PhysRevLett.101.246807,1674-1056-23-3-037304}. The gap, however, exponentially decreases with the growth of the strip width: For $w = 1\,$$\mu$m [see Fig.~\ref{fig:fig2}(a)] the gap is only about 5~$\mu$eV. At $k_y \approx \pm \gamma/A$ the dispersion of the bulk states has pronounced extrema 
originating from the light-hole--heavy-hole mixing [c.f. Fig.~\ref{fig:zero_B}(a)]. According to Eq.~\eqref{eq:velocity}, such a shape of the bulk energy spectrum leads to a flattening of the edge-state dispersion curves. For the dispersion of the edge states shown in Fig.~\ref{fig:fig2} we find the effective velocity $v \approx 2.4\times10^6$\,cm/s whereas calculations with the same parameters but $\gamma = 0$ (not shown) yield $v \approx 5.8\times10^6$\,cm/s.

\subsection{Energy gap in the edge-state spectrum induced by an in-plane magnetic field}\label{sec:oscillate}

\begin{figure}[t]
\includegraphics[width=0.48\textwidth]{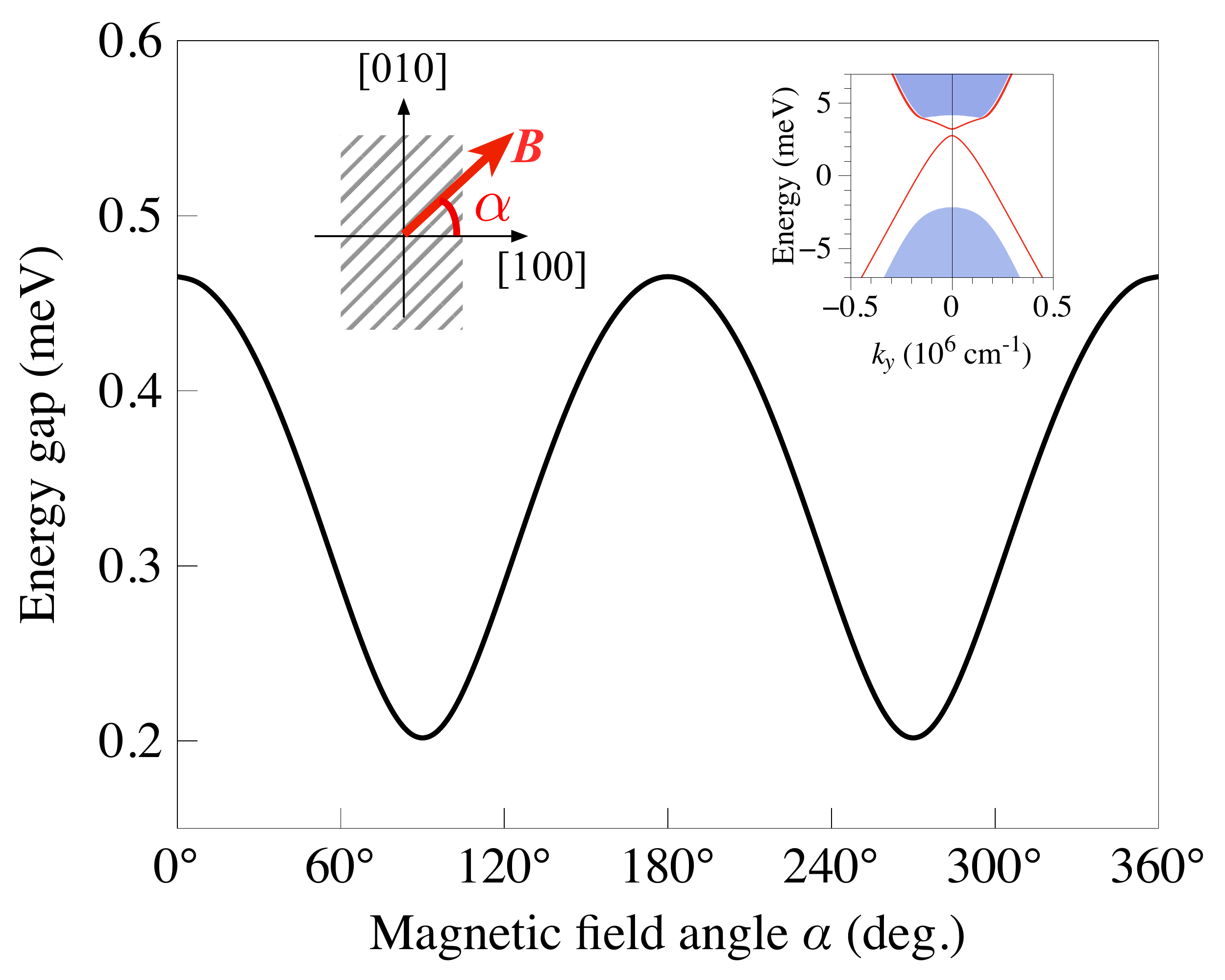}
\caption{\label{fig:fig3}
Energy gap in the edge-state spectrum as a function of the angle $\alpha$ between the in-plane magnetic field $\bm B$ and the edge normal 
calculated for $\delta = -4$~meV and $B = 3$~T. The insets illustrate the geometry under study and the energy spectrum at $\alpha = 0$.}
\end{figure}

As it was shown in Secs.~\ref{sec:B_inplane} and~\ref{sec:orient}, in-plane magnetic field opens a gap in the energy spectrum of the edge states. Due to the light-hole--heavy-hole mixing, the gap depends on the magnetic field direction and the edge orientation. 
Numeric calculation of the gap behaviour upon variation of the magnetic field direction is presented in Fig.~\ref{fig:fig3} for the edges along [010].
The calculated dependence is perfectly described by Eq.~\eqref{eq:gap2} with the effective $g$-factors $|g_{xx}^{(\rm fit)}| \approx 2.7$, $|g_{yy}^{(\rm fit)}| \approx 1.2$. However, the obtained $g$-factors are far smaller than those estimated from Eq.~\eqref{eq:gfactors}, $|g_{xx}| \approx 10.2$, $|g_{yy}| \approx 6.2$, indicating that the ``electron-hole'' asymmetry (${\cal D} \neq 0$) neglected in the analytical theory considerably affects the Zeeman splitting. The value of the $g$-factor anisotropy $(|g_{xx}^{\rm fit}| - |g_{yy}^{\rm fit}|)/ (|g_{xx}^{\rm fit}| + |g_{yy}^{\rm fit}|) \approx 0.38$ is in a much better agreement with the analytical theory which yields $(|g_{xx}| - |g_{yy}|)/(|g_{xx}| + |g_{yy}|) \approx 0.24$.

\begin{figure}[t]
\includegraphics[width=0.48\textwidth]{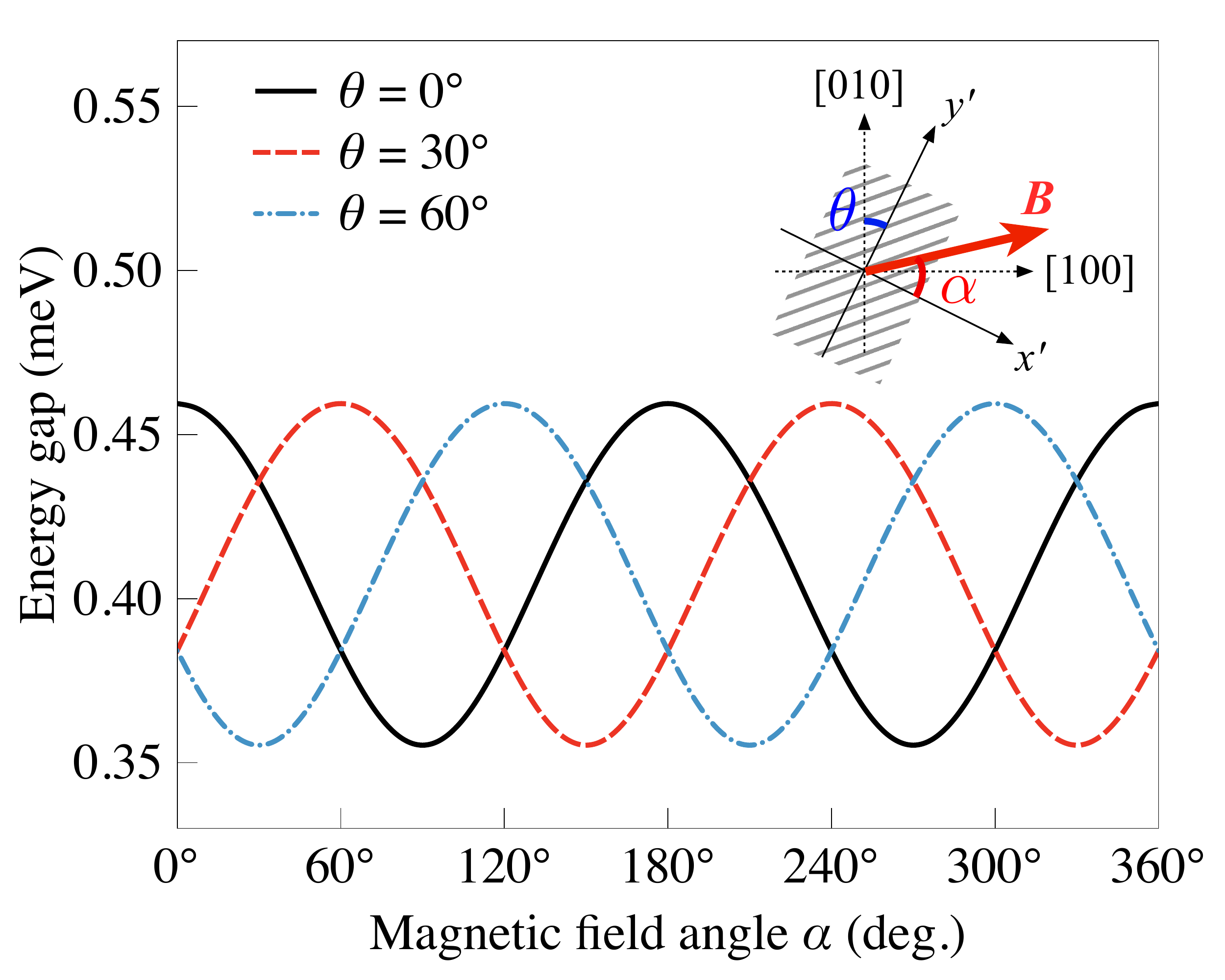}
\caption{\label{fig:fig4}
Energy gap in the edge-state spectrum as a function of the angle $\alpha$ between the in-plane magnetic field $\bm B$ and the edge normal for different orientations of the strip edges. The curves are calculated for $\delta = -10$~meV and $B = 3$~T. The inset illustrates the geometry under study.}
\end{figure}

Figure~\ref{fig:fig4} presents the calculated edge-state energy gap as a function of the angle $\alpha$ for different crystallographic orientations of the strip edges. In agreement with Eq.~\eqref{eq:gap_theta}, the angular dependence of the gap acquires a phase $2\theta$ for the edge tilted by the angle $\theta$ with respect to the $[010]$ axis.

\subsection{Out-of-plane magnetic field}\label{sec:landau}

\begin{figure}[b]
\includegraphics[width=0.48\textwidth]{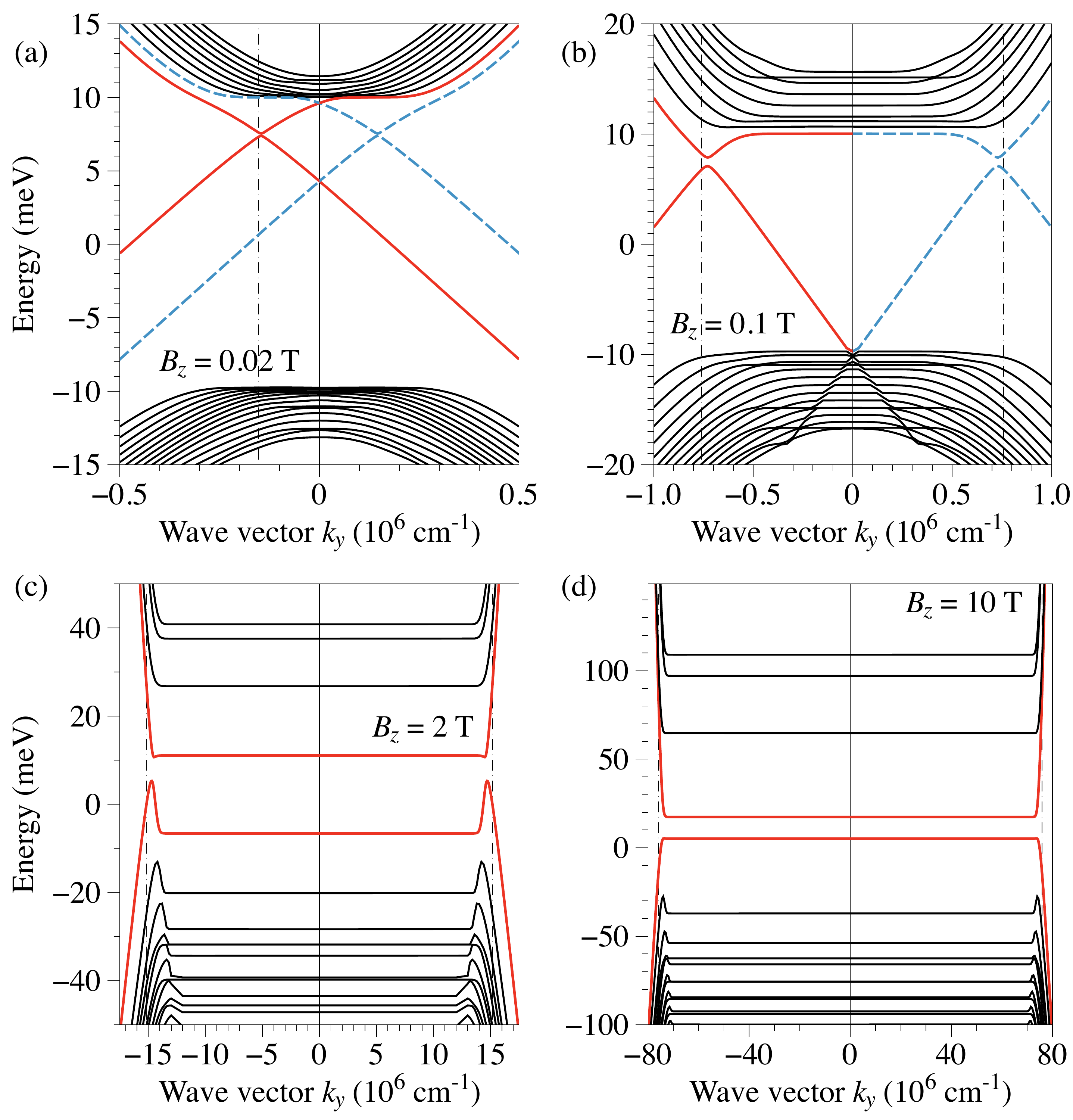}
\caption{\label{fig:fig5}
Electron energy spectra of the strip subject to out-of-plane magnetic field $B_z$.
The spectra are calculated for $w = 1$~$\mu$m and $\delta = -10$~meV. Solid and dashed dispersion curves in panels (a) and (b) correspond to the states localized at the left and right edges, respectively. Dashed-dotted vertical lines indicate the positions $k_y = \pm w/2l_B^2$ for each value of magnetic field.}
\end{figure}

\begin{figure}[htpb]
\includegraphics[width=0.48\textwidth]{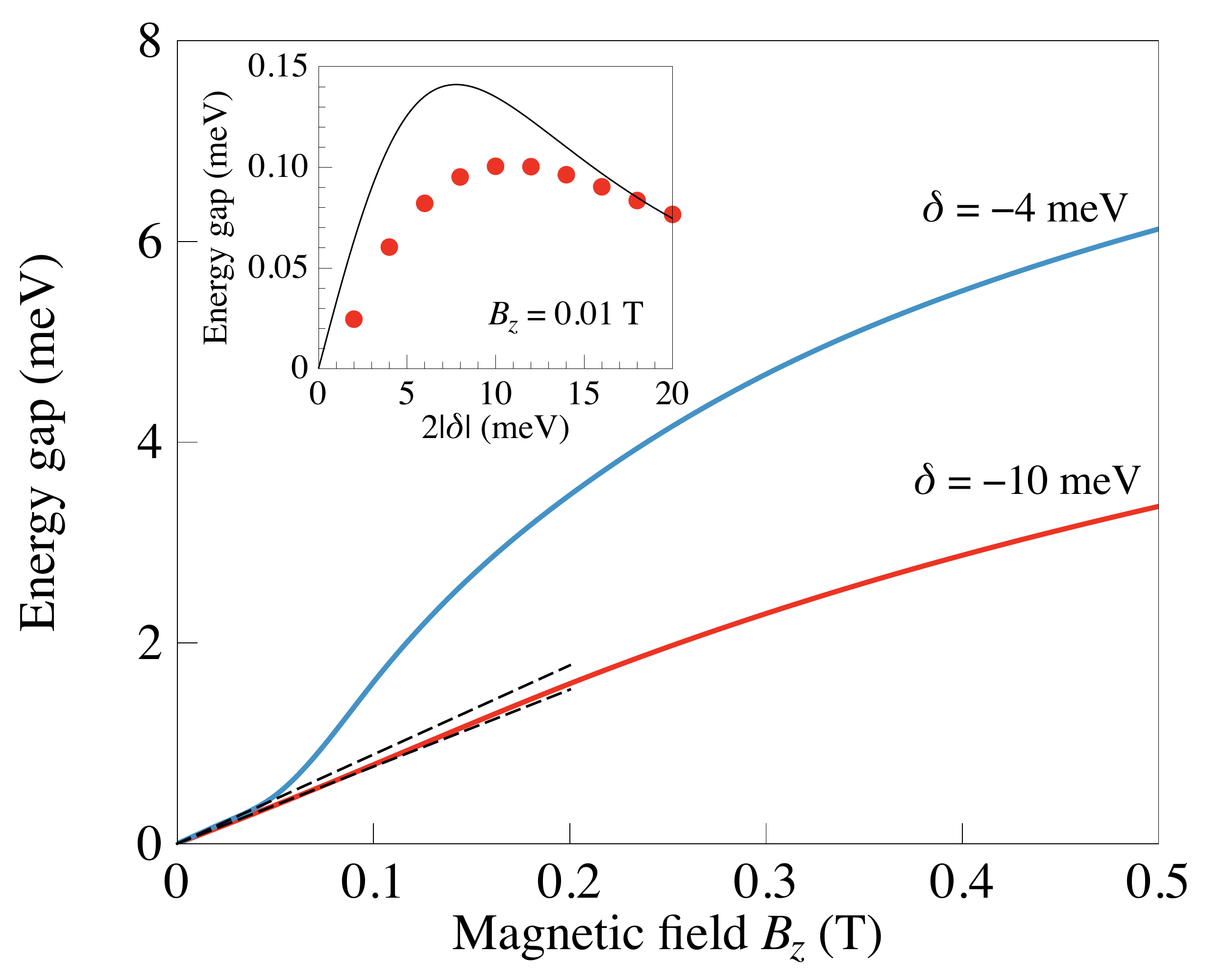}
\caption{\label{fig:fig6}
Energy gap in the edge-state spectrum of a strip as a function of the out-of-plane magnetic field. Dashed lines depict the linear behaviour in the region of small fields. The inset shows the gap at $B_z = 0.01$~T as a function of the bulk gap $2|\delta|$, dots present the results of numerical calculations, solid curve shows the analytical dependence at $\mathcal D = 0$ calculated after Eq.~\eqref{eq:gap_Bz}.
}
\end{figure}

Figure~\ref{fig:fig5} presents electron energy spectra of the strip subjected to an out-of-plane magnetic field at four increasing fields $B_z = 0.02$~T, $0.1$~T, $2$~T, and $10$~T. The spectra are calculated for the bulk gap $2|\delta| = 20$~meV. In accordance with the results of Sec.~\ref{sec:B_outplane}, the out-of-plane magnetic field opens the gap in the edge-state spectrum. The dependence of the edge-state gap $\varepsilon_{\rm gap}$ on $B_z$ is shown in Fig.~\ref{fig:fig6}. At small magnetic fields, this dependence is linear, and for a given magnetic field, $\varepsilon_{\rm gap}$ is a non-monotonic function of the bulk gap $2|\delta|$ (see the inset in Fig.~\ref{fig:fig6}), in agreement with Eq.~\eqref{eq:gap_Bz}.    

The out-of-plane magnetic also results in diamagnetic shifts of the dispersion curves, corresponding to the states localized at the opposite edges. These shifts along $k_y$ are clearly seen at small magnetic field, see Figs.~\ref{fig:fig5}(a) and (b). We note that the relative shift of the left-edge and right-edge energy spectra is independent of the magnetic field gauge 
and, for a wide enough strip, is given by $w/l_B^2$. In our calculations we use the gauge with $x_c=0$ resulting in the symmetric diamagnetic shifts $\pm w/2l_B^2$ with respect to $k_y = 0$.

At large magnetic fields, Figs.~\ref{fig:fig5}(c) and (d), the energy levels become flat for all $k_y$ (except the narrow ranges around
$k_y = \pm w/2l_B^2$), which corresponds to the formation of bulk Landau levels. In particular, two levels in the vicinity of $\eps = 0$ are the bulk zero modes with the energies $\eps_0^{(\pm)}$ given by Eq.~\eqref{eq:zero_modes}. The highly dispersive states at $k_y = \pm w/2l_B^2$ are localized at the strip edges and correspond to the electron and hole chiral edge modes. Thus, the out-of-plane magnetic field finally drives the system into the quantum Hall effect phase.

\section{Conclusions} \label{sec:concl}

To conclude, we have presented the microscopic theory of the electron structure and the Zeeman effect for helical edge states emerging in two-dimensional topological insulators based on HgTe/HgCdTe quantum wells. The theory takes into account strong natural interface inversion asymmetry of the (001)-grown HgTe/HgCdTe quantum wells which reflects the real spatial symmetry described by the $D_{2d}$ point group. The interface inversion asymmetry leads to the mixing of the electron-like and heavy-hole subbands forming the helical edge states. The subband mixing, described by a single parameter $\gamma$, modifies the edge-state dispersion and leads to spatial oscillations of the edge-state wave functions with the wave length controlled by $\gamma$. External magnetic field applied to the quantum well structure destroys the topological protection of the helical states and opens the gap in the edge-state spectrum. For the in-plane magnetic field, the subband mixing gives rise to a strong anisotropy of the edge-state effective $g$-factor which also becomes dependent on the crystallographic orientation of the edge. The $g$-factor anisotropy results, in turn, in variation of the edge-state gap with magnetic field direction and edge orientation. Weak magnetic field normal to the quantum well plane couples the counter propagating edge states and opens the gap in the edge-state spectrum due to the subband mixing, whereas strong normal magnetic field drives the system into the phase of the quantum Hall effect with the formation of chiral electron and hole edge states.

\acknowledgements

The work was partly supported by the Russian Foundation for Basic Research and the ``Dynasty'' Foundation.



\begin{thebibliography}{99}

\bibitem{Bernevig15122006}
B. A. Bernevig, T. L. Hughes, and S.-C. Zhang,   Quantum Spin Hall Effect and Topological Phase Transition in HgTe Quantum Wells, Science \textbf{314}, 1757 (2006).

\bibitem{Konig:2007it}
M. K\"onig, S. Wiedmann, C. Br\"une, A. Roth, H. Buhmann, L. W. Molenkamp, X.-L. Qi, and S.-C. Zhang,   Quantum Spin Hall Insulator State in HgTe Quantum Wells, Science \textbf{318}, 766 (2007).

\bibitem{Gusev2011}
G. M. Gusev, Z. D. Kvon, O. A. Shegai, N. N. Mikhailov, S. A. Dvoretsky, and J. C. Portal,
Transport in disordered two-dimensional topological insulators,
Phys. Rev. B \textbf{84}, 121302(R) (2011).

\bibitem{RevModPhys.82.3045}
M. Z. Hasan and C. L. Kane,   Colloquium, Rev. Mod. Phys. \textbf{82}, 3045 (2010).

\bibitem{RevModPhys.83.1057}
X.-L. Qi and S.-C. Zhang,   Topological insulators and superconductors, Rev. Mod. Phys. \textbf{83}, 1057 (2011).

\bibitem{Konig:2008fk}
M. K\"onig, H. Buhmann, L. W. Molenkamp, T. Hughes, C.-X. Liu, X.-L. Qi, and S.-C. Zhang,   The Quantum Spin Hall Effect: Theory and Experiment, J. Phys. Soc. Jpn. \textbf{77}, 031007 (2008).

\bibitem{PhysRevLett.101.246807}
B. Zhou, H.-Z. Lu, R.-L. Chu, S.-Q. Shen, and Q. Niu,   Finite Size Effects on Helical Edge States in a Quantum Spin-Hall System, Phys. Rev. Lett. \textbf{101}, 246807 (2008).

\bibitem{PhysRevB.82.113307}
E. B. Sonin. Edge accumulation and currents of moment in two-dimensional topological insulators, Phys. Rev. B \textbf{82}, 113307 (2010).

\bibitem{PhysRevB.91.035310}
P. C. Klipstein,   Structure of the quantum spin Hall states in HgTe/CdTe and InAs/GaSb/AlSb quantum wells, Phys. Rev. B \textbf{91}, 035310 (2015).

\bibitem{enaldiev2015}
V. V. Enaldiev, I. V. Zagorodnev, and V. A. Volkov,   Boundary conditions and surface state spectra in topological insulators, Pis'ma Zh. Eksp. Teor. Fiz. \textbf{101}, 94 (2015).


\bibitem{PhysRevB.85.045310}
O. E. Raichev,   Effective Hamiltonian, energy spectrum, and phase transition induced by in-plane magnetic field in symmetric HgTe quantum wells, Phys. Rev. B \textbf{85}, 045310 (2012).

\bibitem{Cheng:2014ve}
F. Cheng, L. Z. Lin, and D. Zhang,   Transport in a quantum spin Hall bar: Effect of in-plane magnetic field, Solid State Commun. \textbf{188}, 45 (2014).

\bibitem{PhysRevLett.104.166803}
G. Tkachov and E. M. Hankiewicz,   Ballistic Quantum Spin Hall State and Enhanced Edge Backscattering in Strong Magnetic Fields, Phys. Rev. Lett. \textbf{104}, 166803 (2010).

\bibitem{PhysRevB.86.075418}
B. Scharf, A. Matos-Abiague, and J. Fabian,   Magnetic properties of HgTe quantum wells, Phys. Rev. B \textbf{86}, 075418 (2012).

\bibitem{PhysRevB.85.125401}
J.-C. Chen, J. Wang, and Q.-F. Sun,   Effect of magnetic field on electron transport in HgTe/CdTe quantum wells: Numerical analysis, Phys. Rev. B \textbf{85}, 125401 (2012).

\bibitem{PhysRevB.91.235433}
B. Scharf, A. Matos-Abiague, I. \ifmmode \check{Z}\else \v{Z}\fi{}uti\ifmmode \acute{c}\else \'{c}\fi{}, and J. Fabian,   Probing topological transitions in HgTe/CdTe quantum wells by magneto-optical measurements, Phys. Rev. B \textbf{91}, 235433 (2015).

\bibitem{PhysRevB.91.081302}
S. A. Tarasenko, M. V. Durnev, M. O. Nestoklon, E. L. Ivchenko, J.-W. Luo, and A. Zunger,  Split Dirac cones in HgTe/CdTe quantum wells due to symmetry-enforced level anticrossing at interfaces, Phys. Rev. B \textbf{91}, 081302 (2015).

\bibitem{Nanostructures2014}
S. A. Tarasenko, M. O. Nestoklon, M. V. Durnev, and E. L. Ivchenko, Zero-field splitting of Dirac cones in HgTe/CdHgTe quantum wells, Proc. of 22nd Int. Symp. ``Nanostructures: Physics and Technology'' (23-27 June, Saint-Petersburg, Russia), pp. 174-175 (2014).

\bibitem{Olbrich2013} P. Olbrich, C. Zoth, P. Vierling, K.-M. Dantscher, G. V. Budkin, S. A. Tarasenko, V. V. Bel'kov, 
D. A. Kozlov, Z. D. Kvon, N. N. Mikhailov, S. A. Dvoretsky, and S. D. Ganichev, Giant photocurrents in a Dirac fermion system at cyclotron resonance, Phys. Rev. B \textbf{87}, 235439 (2013).

\bibitem{Dantscher2015} K.-M. Dantscher, D. A. Kozlov, P. Olbrich, C. Zoth, P. Faltermeier, M. Lindner, G. V. Budkin, S. A. Tarasenko, V. V. Bel'kov, Z. D. Kvon, N. N. Mikhailov, S. A. Dvoretsky, D. Weiss, B. Jenichen, and S. D. Ganichev, Cyclotron-resonance-assisted photocurrents in surface states of a three-dimensional topological insulator based on a strained high-mobility HgTe film, Phys. Rev. B \textbf{92}, 165314 (2015).

\bibitem{Gerchikov1989}
L. G. Gerchikov and A. V. Subashiev, Non-monotonic behaviour of the energy gap in the film made of a gapless semiconductor,   Fiz. Tekh. Poluprovodn. \textbf{23}, 2210 (1989) [Sov. Phys. Semicond. \textbf{23}, 1368 (1989)].

\bibitem{Winkler20122096}
R. Winkler, L. Y. Wang, Y. H. Lin, and C. S. Chu,  Robust level coincidences in the subband structure of quasi-2D systems, Solid State Commun. \textbf{152}, 2096 (2012).


\bibitem{Mar99}
X. Marie, T. Amand, P. Le Jeune, M. Paillard, P. Renucci, L. E. Golub, V. D. Dymnikov, and E. L. Ivchenko,   Hole spin quantum beats in quantum-well structures, Phys. Rev. B \textbf{60}, 5811 (1999).

\bibitem{Rashba1960}
E. I. Rashba.   Properties of semiconductors with an extremum loop. I. Cyclotron and comninational resonance in a magnetic field perpendicular to the plane of the loop, Fiz. Tverd. Tela \textbf{2}, 1874 (1964).


\bibitem{PhysRevB.83.115307}
M. Orlita, K. Masztalerz, C. Faugeras, M. Potemski, E. G. Novik, C. Br\"une, H. Buhmann, and L. W. Molenkamp,   Fine structure of zero-mode Landau levels in HgTe/Hg$_x$Cd$_{1-x}$Te quantum wells, Phys. Rev. B \textbf{83}, 115307 (2011).


\bibitem{PhysRevB.86.205420}
M. Zholudev, F. Teppe, M. Orlita, C. Consejo, J. Torres, N. Dyakonova, M. Czapkiewicz, J. Wr\'obel, G. Grabecki, N. Mikhailov, S. Dvoretskii, A. Ikonnikov, K. Spirin, V. Aleshkin, V. Gavrilenko, and W. Knap, Magnetospectroscopy of two-dimensional HgTe-based topological insulators around the critical thickness, Phys. Rev. B \textbf{86}, 205420 (2012).


\bibitem{Buttner:2011ve}
B. Buttner, C. X. Liu, G. Tkachov, E. G. Novik, C. Brune, H. Buhmann, E. M. Hankiewicz, P. Recher, B. Trauzettel, S. C. Zhang, and L. W. Molenkamp,   Single valley Dirac fermions in zero-gap HgTe quantum wells, Nat. Phys. \textbf{7}, 418 (2011).

\bibitem{1674-1056-23-3-037304}
C. Zhi and Z. Bin,   Finite size effects on helical edge states in HgTe quantum wells with the spin-orbit coupling due to bulk- and structure-inversion asymmetries, Chin. Phys. B \textbf{23}, 037304 (2014).

\end{thebibliography}

\end{document}